\documentclass[useAMS,usenatbib]{mn2e}
\usepackage{graphicx,graphics,amsmath}

\newcommand{\Om}{\Omega_{\rm M}} 
\newcommand{\half}{\frac{1}{2}} 
\newcommand{\T}{\rm T} 
\newcommand{\Ox}{\Omega_{\rm X}}
 
\newcommand{\Ol}{\Omega_{\Lambda}} 
\newcommand{\bx}{\mathbf x}
\begin{document}
\title {Direction Dependence in Supernova Data: Constraining Isotropy}

\author[Shashikant Gupta and Tarun Deep Saini]
{Shashikant Gupta $^{1,2,3,4}$ and Tarun Deep Saini $^{1,5}$\\
  $^{1}$ Indian Institute of Science, Bangalore, Karnataka, India, 560 012 \\
   $^{2}$ Raman Research Institute, Sadashivanagar, Bangalore, Karnataka, India, 560 080 \\
  $^{3}$ shashikant@physics.iisc.ernet.in\\ 
  $^{4}$ shashikant@rri.res.in \\
  $^{5}$ tarun@physics.iisc.ernet.in \\
} \maketitle

\begin{abstract}
  We revise and extend the extreme value statistic, introduced in
  \cite{gup08}, to study directional dependence in the high redshift
  supernova data; arising either from departures from the cosmological
  principle or due to direction dependent statistical systematics in
  the data. We introduce a likelihood function that analytically
  marginalises over the Hubble constant, and use it to extend our
  previous statistic. We also introduce a new statistic that is
  sensitive to direction dependence arising from living off-centre
  inside a large void, as well as previously mentioned reasons for
  anisotropy. We show that for large data sets this statistic has a
  limiting form that can be computed analytically.  We apply our
  statistics to the gold data sets from \cite{rie04} and \cite{rie07},
  as in our previous work. Our revision and extension of previous
  statistic shows that 1) the effect of marginalsing over Hubble
  constant instead of using its best fit value has only a marginal
  effect on our results. However, correction of errors in our previous
  work reduce the level of non-Gaussianity in the 2004 gold data that
  was found in our earlier work.  The revised results for the 2007
  gold data show that the data is consistent with isotropy and
  Gaussianity. Our second statistic confirms these results.
\end{abstract}

\section{Introduction}

Observations of high-redshift supernovae of Type~Ia
\citep{per99,rie98,rie02,rie04}, along with observations of the cosmic
microwave background \citep{ben02,pag06}, indicate a flat universe
with an accelerating expansion. The existing evidence is
consistent with a universe that is dominated today by a cosmological
constant term in the Einstein's theory of gravity, however,
due to fine tuning required in this model, several alternative
explanations have been suggested \citep{silv09,frie08,sahn06}. If we
phenomenologically treat this unknown energy component as an ideal
fluid --- called the dark energy --- then the pressure of this fluid has
to be negative, that is, for an equation of state $p=w\rho$, we have
$\rho > 0$ and $w<0$. In this description the cosmological constant
model has an equation of state
given by $p=-\rho$, and thus we expect that a phenomenological
modeling of the cosmological data with a constant $w$ should give $w
\simeq -1$, which turns out to be the case; the current data 
indicating one-third of the total density in the form of dark and
baryonic matter and two-thirds in the form of the cosmological
constant --- a model that is known as the $\Lambda$CDM model. However,
it should be noted that the equation of state for the dark energy
could differ from $w=-1$ in the past when the dark energy is expected
to be subdominant (for consistency with constraints arising from the
observations of the microwave background). Indeed, in many plausible
models of dark energy the equation of state approaches $w=-1$ only in
the recent past and thus may show only tiny departures from $w=-1$ at
low redshifts, while at high redshifts it might remain subdominant and
therefore would have a weaker observational signature. Although it is
perhaps hard to extend supernova (SN) observations beyond $z \simeq 3$
or so, neutral hydrogen observations of the post-reionization epoch
may be able to provide constraints on dark energy all the way up to $z
\simeq 6$ \citep{bhar09}.

To resolve this issue we require data that is precise enough to
discern tiny variations in the dark energy. It is also required that
data be available at a large number of redshifts to constrain the
detailed temporal behaviour of dark energy. At present the only data
that comes reasonably close to these requirements is provided by the
observations of the high redshift type Ia supernovae (SNe), which are
believed to be standard candles --- a claim that may be doubtful
considering the relatively poorly understood physics, and the
possibility of physical mechanisms such as dust in the inter-galactic
medium that systematically dims them. Another concern is that the SN
data sets are usually collated from several different experiments that
might have slightly different systematics, due either to instrumental
effects or the fact that they observe different directions in the sky;
for example systematic errors in correcting Galactic or source galaxy
dust extinction might leave residual anisotropies in data. These
considerations imply that to obtain precise information about the
behaviour of dark energy we should first have a good knowledge of the
statistical properties of SNe, both random as well as systematic.

The standard model of cosmology is based on the Cosmological Principle
\citep{pee93}, which states that the Universe is homogeneous and
isotropic on large scales.  Recent work provides some evidence for
what is known as the Hubble Bubble \citep{zeh98,jha07} indicating that
we might be living inside a large void that has a different value of
the Hubble constant inside from what is outside the bubble. There is
evidence for such large scale voids in the CMB maps as well
\citep{cruz05,olv06,cruz06,cruz07}, suggesting that such large voids
are not implausible. There is no reason for us to believe that we are
living at the precise centre of such a void. Therefore, if we happen to be living in such a void
then various distance measures, such as the luminosity
distance and the angular diameter distance, may not be
isotropic. \cite{gup08} (hereafter GSL08) used the extreme value
statistics to show that the two SN data sets,
\cite{rie04} (GD04) and \cite{rie07} (GD07), do show some evidence for
direction dependence. Several other works have also indicated
either systematic problems with the high-redshift SN data or
directional dependence in the SN data and other probes
\citep{kol01,ness04,ness05,ness07, jain106,jain206,coo09}. 

We recently found that our analysis in GSL08 contained mistakes due to
a coding error.  In this paper our main task is 1) to provide revised
results and to extend the previous work by introducing a statistic
that marginalises over the Hubble parameter instead of using its best
fit value 2) to introduce a new statistic that has a greater
sensitivity to the signatures of living off-centre inside a large
void. We also give asymptotic form for this statistic, which makes it
easier to use it for large data sets.

The plan of the paper is as follows. In \S~2 we introduce the
likelihood function marginalised over the Hubble constant and provide
a comparison with what we obtain using the Bayesian statistic. In \S~3
we introduce the two statistics. In \S~4 we present our results and
\S~5 we end with conclusions.

\section{Marginalization over $H_0$ and $M$}
For a given SN the measured quantity, the distance modulus $\mu$, is
the difference between the apparent and the absolute magnitude
\begin{equation}
 \mu(z) = m(z) - M\,, 
\label{eq:mu1}
\end{equation}
where the apparent magnitude $m(z)$ is a function of the intrinsic
luminosity of a SN, the redshift $z$ and the cosmological parameters;
and $M$ is the absolute magnitude of a type~Ia SN. The distance
modulus can be expressed in terms of the luminosity distance $D_L$ as
\begin{equation}
\mu(z) = 5 \log \left[{D_L(z)/{\rm Mpc}} \right] + 25\,,
\label{eq:mu}
\end{equation}
where the luminosity distance is given by
\begin{equation}
D_L(z) = \frac{c (1+z)}{H_0}\int_{0}^{z} \frac{dx}{h(x)}\,,
\label{eq:D_L}
\end{equation}
where $h(z; \Om,\Ox)=H(z; \Om,\Ox)/H_0$, and thus depends only on the
cosmological parameters; the matter density $\Om$ and the dark energy
density $\Ox$. The prescription for the variation of
dark energy with redshift has to be specified separately. For example, in the
$\Lambda$CDM model the energy density in the dark energy $\Ol$ remains a
constant. In Eq~\ref{eq:mu1} the dependence of the measured quantity
$\mu$ on $M$ is linear. Since $\mu$ depends on the logarithm of the
luminosity distance it is clear that $\mu$  depends linearly on the logarithm of
the Hubble parameter $H_0$. Usually the data are given in terms of
Eq~\ref{eq:mu}, where the constant
$M$ has already been marginalised over. Thus, instead of two
nuisance parameters we are left with only one --- the Hubble
constant $H_0$. 

The Hubble parameter could in principle be measured using the
observations of the low-redshift distance-redshift relationship,
however, the quantities of real interest are the cosmological
parameter $\Om$ and $\Ox$. Thus it would be useful to marginalise over
the nuisance parameter $H_0$. Although this can be done numerically
while estimating the cosmological parameters, it turns out that our
statistic in \S~3.2 does not allow this. It is clear that the Hubble
constant term can be eliminated by considering the difference of two
magnitudes
\begin{equation}
\label{eq:diffmu}
\mu_i - \mu_j = 5 \log \left[\frac{ (1+z_i) \int_{0}^{z_i} dx/h(x)}{ (1+ z_j)\int_{0}^{z_j} dx/h(x)} \right]\,.
\end{equation}
If the error on $\mu_i$ is
$\sigma_i$ then the standard error for $\mu_i - \mu_j$ is given by
$\sigma = \sqrt{\sigma^2_i +
\sigma^2_j}$, where we have assumed that the errors on the 
two magnitudes are statistically independent. If we have a large
number, $N$, of SNe in the sample then dividing it into two equal
halves we can form $N/2$ such differences (assuming an even $N$), and
estimate parameters that would be independent of $H_0$. However, the
degrees of freedom reduce by half in this process, thereby reducing
the amount of information that can be extracted from data. Another
reason for our not choosing this method is the fact that we are
interested in quantifying direction dependence in data and therefore
our statistics depend on the direction each SN is
observed. Specifically, we consider statistics that depend on the
difference of quantities computed on the opposite hemispheres and then
maximizing this difference over all the directions. It is clear that
such a procedure does not allow the use of above method.

For marginalisation we instead use a method that is based on
subtracting the magnitude of an arbitrarily chosen low-redshift (so
its magnitude depends only on the Hubble constant) anchor SN
from the magnitudes of SNe in a data set and then marginalising over
magnitude of anchor SN. The resulting likelihood function is derived
in the Appendix. This is equivalent to
marginalising over the nuisance parameter $H_0$ with a Gaussian prior
centered around its value derived from the anchor SN, alongwith the
corresponding standard deviation. The method can be easily generalized
to a case where the priors on $H_0$ are specified separately as
described in the Appendix.

\subsection{Comparison with Bayesian Marginalization} 

\begin{figure}
\centering
\includegraphics[width=0.50\textwidth]{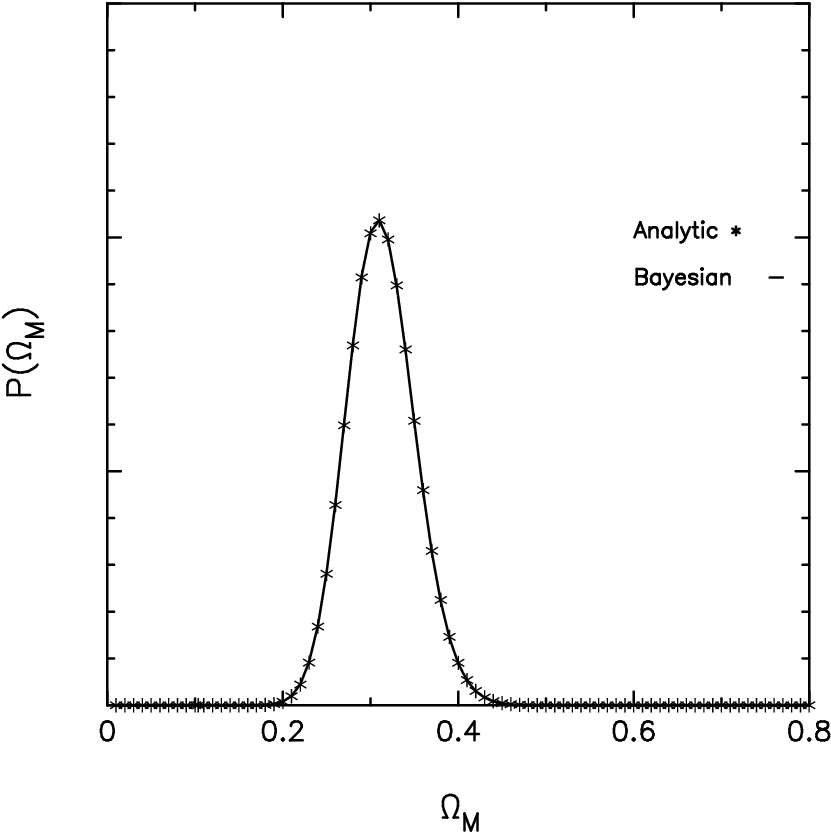}
\caption{Comparison of numerically marginalised probability density for
the $\Om$ with the analytic marginalisation according to the
likelihood function in Eq~\ref{eq:marglik}. The two are
indistinguishable.  }
\label{fig:comparison}
\end{figure}

We now compare the analytically marginalised likelihood function in
Eq~\ref{eq:marglik} to the results of Bayesian marginalisation over
$H_0$. For this purpose we consider parameter estimation for GD04. We
assume a flat $\Lambda$CDM universe for this exercise. The
dimensionless Hubble parameter is given by
\begin{equation}
h(z) = \sqrt{\Om (1+z)^3 + \Ol}\,,
\end{equation}
and flatness implies $\Ol = 1 - \Om$. Thus, the only free parameters
are $H_0$ and $\Om$. The normalized likelihood function is given by
\begin{equation}
L(\mathbf x | \Om, H_0) = \frac{1}{(2\pi)^{N/2}\sqrt{ \det {\mathbf \Sigma}}}\exp \left (
-\half \mathbf{ x}^{\T} \mathbf{\Sigma}^{-1} \mathbf{x} \right )\,.
\end{equation}
Here $x_i = \mu_i - \mu_{\rm theory} (z_i; H_0,\Om)$, where the
subscript $i=1,\ldots,N$ and $\Sigma_{ij} = \delta_{ij}\,\sigma_i^2$ is the
covariance matrix. The superscript 'T' denotes the matrix operation of
taking the transpose of a matrix.  The posterior probability for
parameters $\Om$ and $H_0$ is given by
\begin{equation}
P(\Om,H_0| \mu_i) = \frac{L( \bx |\Om,H_0)P(H_0,\Om)}{P(\bx)}\,.
\label{eq:bayes}
\end{equation}
We choose a uniform prior for $\Om$ in the range $0 \le \Om \le 1$ and
for $H_0 = 100h\,\rm km \,s^{-1}\,Mpc^{-1}$ in the range $0.4 \le h \le 1$. 
The marginalised probability distribution for the matter density is given by
\begin{equation}
P(\Om| \mu_i) \propto \int_{H_0} L( \bx |\Om,H_0)P(H_0,\Om) dH_0\,.
\end{equation}
The probability density is normalized after carrying out the
integration. A similar probability density function for the matter
density can be obtained by the Bayesian inversion of the marginalised
likelihood function given in Eq.~A5. Figure~\ref{fig:comparison} plots a
comparison between the two probability densities and shows
that the two distributions are nearly identical.

\begin{figure}
\centering
\includegraphics[width=0.50\textwidth]{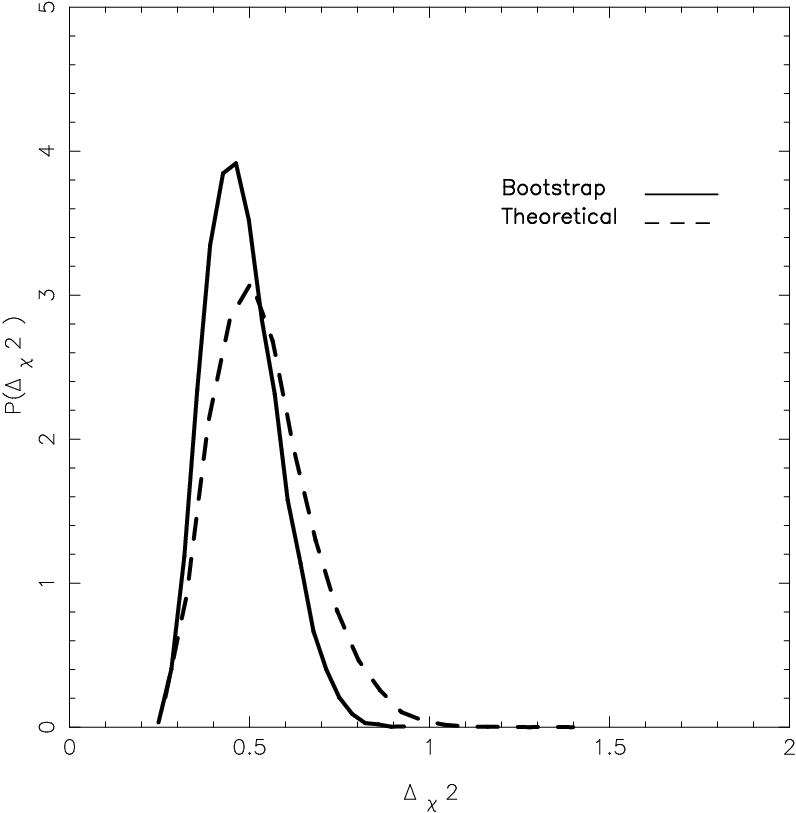}
\caption{A comparison of theoretical and bootstrap probability distributions for simulated data. 
The data comprises $157$ SNe, whose positions on the sky were
generated randomly. There was a mistake in a similar figure in GSL08
where the discrepancy between these distributions was larger.  }
\label{fig:sim}
\end{figure}

\section{Two Statistics} 
\label{sec:method}
For completeness some details are repeated here from GSL08. For our
analysis we have considered a flat $\Lambda$CDM universe, which can be
easily generalized to a more general model of dark energy.

\subsection{$\Delta_{\chi^2}$ statistic}

We consider subsets of the full data set to construct our statistic
comprising  $N_{\rm subset}$ data points. For analytic  marginalisation  we keep two anchor
SNe aside. 
Since the $\Lambda$CDM model fits the gold data sets GD04 and GD07 well,
we first obtain the best fit to the \emph{full gold
data sets} using the marginalised likelihood function given in  Eq~\ref{eq:marglik}, 
and then for each SN we calculate the residuals \mbox{ $\chi_i = [\mu_i -
\mu^{\Lambda\rm CDM} (z_i;\Om)] - [\mu_0 -
\mu^{\Lambda\rm CDM} (z_0;\Om)] $}, which is free of $H_0$ by virtue of Eq~\ref{eq:diffmu}. 
The standard error for a SN at redshift $z_i$ is $\sigma_{\mu}(z_i)$, and we assume that
the errors on SNe are statistically uncorrelated.

We define $\chi^2_R = \chi^2_M/N_{\rm
subset}$, where the marginalised $\chi^2_M$ is defined in
Eq.~\ref{eq:mchisq}. $\chi^2_R$ indicates the statistical scatter of
the subset from the best fit $\Lambda$CDM model.  Its expectation
value is unity (see Appendix for a proof), that is $\langle \chi^2_R
\rangle = 1$.  We divide the data into two hemispheres labeled by the direction
vector $\hat{n}$, and take the difference of the $\chi_R^2$
computed for the two hemispheres separately to obtain $\Delta
\chi_{\hat{n}}^2 =\chi^2_{R1} - \chi^2_{R2} $, where label '1' corresponds to
that hemisphere towards which the direction vector $\hat{n}$ points
and label '2' refers to the opposite hemisphere. We take the absolute
value of $\Delta
\chi_{\hat{n_i}}^2$ since we are interested in the largest magnitude of this quantity. 
We then vary the direction $\hat{n}$ across the sky to obtain the
maximum absolute difference
\begin{equation}
\Delta_{\chi^2} = {\rm max} \{| \Delta \chi_{\hat{n}}^2 |\}\,\,. 
\end{equation}

As shown in GSL08, the distribution of $\Delta_{\chi^2}$ follows a
simple, two parameter Gumbel distribution, characteristic of extreme
value distribution type~I \citep{ken77, gum65},
\begin{equation}
P(\Delta_{\chi^2}) =\frac{1}{s} \exp \left[ -\frac{ \Delta_{\chi^2}
-m}{s}\right]\,\exp\left [-\exp\left(-\frac{\Delta_{\chi^2}
-m}{s}\right)\right]\,\,,
\end{equation} 
where the position parameter $m$ and the scale parameter $s$
completely determine the distribution. To quantify departures from
isotropy we need to know the theoretical distribution, which is
calculated numerically by simulating several sets of Gaussian
distributed $\chi_i$ on the gold set SN positions and obtaining
$\Delta_{\chi^2}$ from each realization. For comparison with theory we follow GSL08 and compute a
bootstrap distribution by shuffling the data values $z_i$, $\mu(z_i)$
and $\sigma_{\mu}(z_i)$ over the SNe positions (for further details
see GSL08).

As mentioned in GSL08, the above procedure separates the data sets
into hot and cold SNe that have large and small dispersions with
respect to the best fit model. However, note that these two sets could
still indicate the same cosmology, albeit with a different value of
$\chi^2$. To ameliorate this deficiency we now introduce a new
statistic that does not suffer from this artifact.

\subsection{ $\Delta_{\chi}$ statistic}

As mentioned above, $\chi_i^2$ does not contain information about
whether the SN is above or below the fit.  An obvious generalization that does contain information
regarding whether the SN at a redshift is closer or farther
from us can be obtained by considering a statistic based on $\chi_i$s. We consider two subsets of data defined by two hemispheres labeled by the direction vector
$\hat{n}$, containing $N_{\rm north}$ and $N_{\rm south}$ SNe, where the total number of SNe, 
 $N = N_{\rm north} + N_{\rm south}$,  and define the quantity
\begin{equation}
\Delta\chi_{\hat{n}} = \frac{1}{\sqrt{N}} \left( \sum_{i=1}^{N_{\rm north}} \frac{\chi_i}{\sigma_i} - \sum_{j=1}^{N_{\rm south}} \frac{\chi_j}{\sigma_j} \right )\,.
\end{equation}
Clearly $\langle \Delta \chi_{\hat{n}} \rangle = 0$ and $\langle
(\Delta \chi_{\hat{n}})^2 \rangle = 1$. From the central limit theorem
\citep{ken77} it follows that for $N \gg 1$, the quantity
$\Delta\chi$ follows a Gaussian distribution with a zero mean and unit
variance. As in the previous case we maximize this quantity by
varying the direction $\hat{n}$ across the sky to obtain the maximum
absolute difference
\begin{equation}
\Delta_{\chi} = {\rm max} \{| \Delta \chi_{\hat{n}} |\}\,\,. 
\end{equation}

Unlike the $\Delta_{\chi^2}$ statistic this statistic is not
marginalised over the Hubble constant since our results show 
that marginalising over it instead of using its best fit value has only a marginal effect on
$\Delta_{\chi^2}$. Moreover, in the limit $N \gg 1$, and assuming a uniform sky
coverage, we expect the two hemispheres to contain roughly an equal number of
SNe. In this case it is clear that $\Delta_{\chi}$ would depend only weakly
on $H_0$.

This statistic differs from the previous one in that 
the $\Delta_{\chi}$ statistic has
a theoretical limit where the position and the shape parameters can
be determined analytically. Given $N_d$ independent directions on the sky we are
essentially determining the maximum of a sample of size $N_d$ where
the individual numbers are drawn from a Gaussian distribution with a
zero mean and unit variance. In the limit $N_d \gg 1$ the parameters are 
given by \citep{haan06} 
\begin{eqnarray}
m &=& \sqrt{2\log N_d -\log \log N_d - \log 4\pi}\\
s &=& \frac{1}{m}
\end{eqnarray}
where we have to additionally assume that the number of SNe $N\gg 1$,
since the distribution for $\chi$ becomes Gaussian only in this
limit. This is convenient since at least for large data sets, which
will be available in the future, a comparison with theory becomes
simpler. However, for a smaller number of SNe there is a possibility
that not all directions are independent, in fact, it is quite possible
that two directions contain exactly same subsets in the two
hemisphere. In this situation is is clear that the total independent
directions is a smaller number than $N_d$ and thus theoretical
distribution would be rightward shifted and also more sharply
peaked. For this reason we also calculate the bootstrap distribution 
and the theoretical distribution in the same manner as for the previous
statistic.

\section{Results}

\begin{figure}
\centering
\includegraphics[width=0.50\textwidth]{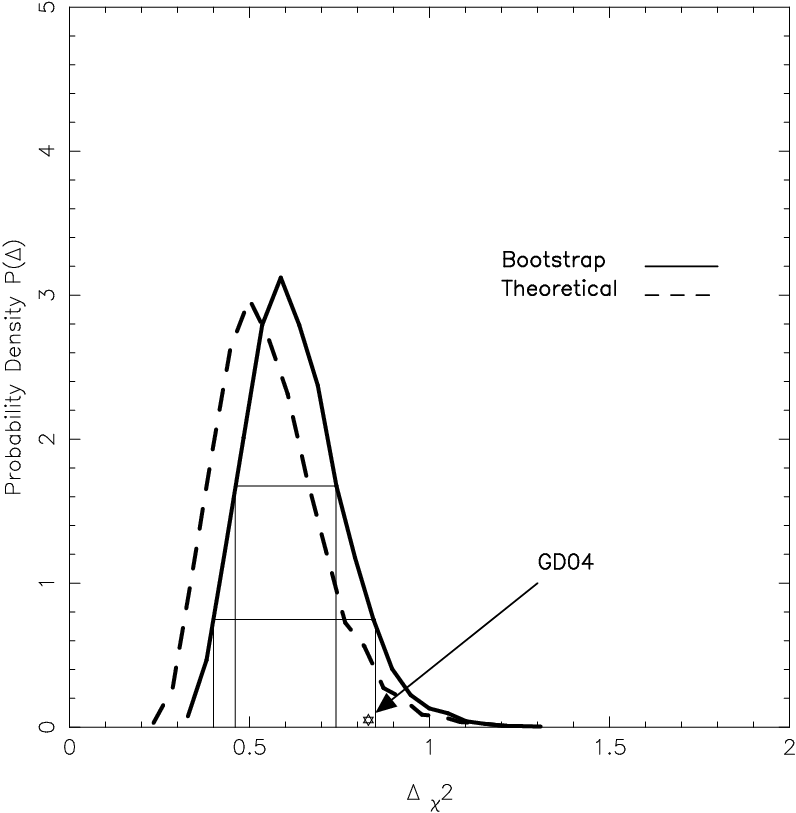}
\caption{The theoretical and the bootstrap probability distributions for 
GD04 for the $\Delta_{\chi^2}$ statistic. Theoretical distribution is
shifted to the left compared to what we find for our simulated data in
Figure~\ref{fig:sim}, which uses Gaussian deviates suggesting evidence
for non-Gaussianity.}
\label{fig:old04}
\end{figure}

\begin{figure}
\centering
\includegraphics[width=0.50\textwidth]{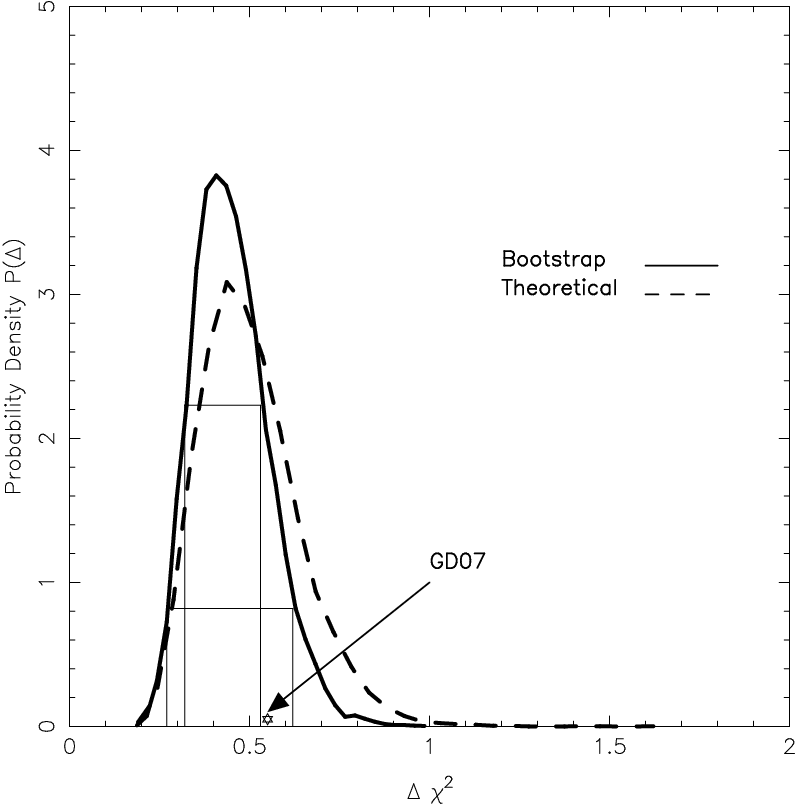}
\caption{The theoretical and the bootstrap probability distributions for GD07 
for the $\Delta_{\chi^2}$ statistic. Theoretical distribution is
compatible with Figure~\ref{fig:sim} for the simulated data, which
implies Gaussianity for the residuals. The $\Delta_{\chi^2}$ for GD07
is within one sigma of the mode and thus the data are consistent with
isotropy.  }
\label{fig:old07}
\end{figure}

In GSL08 we discussed a specific bias in the bootstrap distribution,
showing that it is shifted slightly to the left of the theoretical
distribution due to the fact that theoretical distribution is obtained
by assuming $\chi_i$s to be Gaussian random variates with a zero mean
and unit variance. Theoretical $\chi_i$s are
unbounded, however the bootstrap distribution is obtained by
shuffling through a \emph{specific realization} of $\chi_i$ where the $\chi_i$s 
are obviously bounded. It is clear that on the
average this should produce slightly smaller values of $\Delta_{\chi^2}$ in
comparison to what one expects from a Gaussian distributed
$\chi_i$s. The same problem persists in the Hubble constant
marginalised version of $\Delta_{\chi^2}$. 

However, the analysis in GSL08 had a numerical bug that produced a 
mismatch between the bootstrap and theoretical distributions
that is worse than what one obtains upon correction. 
Therefore, for reference we plot the results for a
new simulation in Figure~\ref{fig:sim} with a total of $157$ SNe. This
may be compared with Figure~3 of GSL08. The discrepancy
has been reduced after correction. Our results in
this paper should be interpreted with respect to
Figure~\ref{fig:sim} in this paper. Concerns regarding the small number of SNe in
data sets and its effect on the efficacy of our method can be
addressed (as in GSL08) by noting that this figure is produced with
only $157$ SNe and the theoretical and the bootstrap distributions look
similar.

\subsection{Results: $\Delta_{\chi^2}$ Statistic}

\begin{table}
\begin{center}
\caption{
  The model parameters for GD04 and GD07 are tabulated  here.}  
  \label{tbl-oldfit}
   \bigskip

\begin{tabular}{ccrrrrr}
\hline
Model & Set & $\Om$   &  $\chi^2$ \\
\hline
$\Lambda$CDM & GD04 & 0.32  &  177.1 \\
$\Lambda$CDM & GD07 & 0.33  &  158.7 \\
\hline\hline                                                                                        
\end{tabular}
\end{center}
\end{table}

\begin{table}
\begin{center}
\caption{
  Direction for maximum $\Delta$ GD04 and GD07 are tabulated  here.}
  \label{tbl-olddel} \bigskip

\begin{tabular}{crrrrrr}
\hline
Model & Set & $\Delta_{\chi^2}$ &  long  &  lat \\
\hline
$\Lambda$CDM & GD04 & 0.83 & 90. &  44.7 \\
$\Lambda$CDM & GD07 & 0.53 & 347. &  27. \\
\hline\hline
\end{tabular}
\end{center}
\end{table}

This statistic is different from the one used in GSL08 in the fact
that here we have marginalised over the Hubble parameter. However, our
results differ significantly from those presented in GSL08 due to the
fact that we have corrected a numerical bug in the calculation used in
GSL08 \footnote{In GSL08 the theoretical and the bootstrap distributions were handled
by different codes, and one of them had a bug thereby 
creating a greater discrepancy between them than should have been.} In 
Table~\ref{tbl-oldfit} we give the best fit values of $\Om$ using
the likelihood function in Eq.~\ref{eq:mchisq} for both data
sets. This is the model that we subtract from data to produce the
residual $\chi_i$s. We note that GD07 gives slightly higher value of
$\Om$ compared to GD04. The direction and the value of
($\Delta_{\chi^2}$) is presented in Table~\ref{tbl-olddel}.

{\bf GD04}: In Fig~\ref{fig:old04} we plot the bootstrap and the
theoretical distribution expected for GD04 and mark the position of
GD04. Comparison with Figure~1 in GSL08 shows that for GD04 the
essential difference is that the theoretical curve has shifted to the
left while the bootstrap distribution is almost identical, suggesting that 
the effect of marginalisation over the Hubble parameter is small. Comparison
with Figure~\ref{fig:comparison} shows that there is still a weak
signature of non-Gaussianity since the theoretical distribution
instead of being to the right of the bootstrap is instead shifted to
the left.

{\bf GD07} Results are plotted in Fig~\ref{fig:old07}. This
should be compared to Figure~2 of GSL08. As in the case of GD04, we
find that the theoretical distribution has shifted to left after
correcting the bug in the code while  the bootstrap distribution
is unchanged.  However, in this case, due to the correction of an additional
mistake in the code, GD07 has shifted its position to the left. 
Comparison with Figure~\ref{fig:sim} shows that our revised results are
compatible with the absence of any features of non-Gaussianity in the
data. GD07 sits at about one sigma from the mode of the
distribution, thus the directional dependence is weaker than
found in GD04.


\subsection{Results: $\Delta_{\chi}$ statistic}

The essential difference between the previous statistic and this one
is that this statistic is sensitive to a SN being above or below the
best fit. As mentioned in \S~3.2, in the limit $N \gg 1$ and $N_d \gg
1$, the distribution for $\Delta_{\chi}$ is fully determined. In the
results below we call this the \emph{analytic} distribution.

Since this statistic is not marginalised over the Hubble constant, as
a first step we fit the data sets to the two parameters $H_0$ and
$\Om$ in order to obtain the residuals $\chi_i$. We tabulate the
best-fit value of these parameters in Table~\ref{tbl-newfit}.  The
value of $\Delta_{\chi}$ and the direction in which the maximum occurs
are tabulated in Table~\ref{tbl-newdel}. The quoted directions 
in both cases refer to a negative value $\Delta_{\chi}$, thus in both cases
the direction points in the direction where SNe appear closer to us than the best fit.

\begin{figure}
\centering
\includegraphics[width=0.50\textwidth]{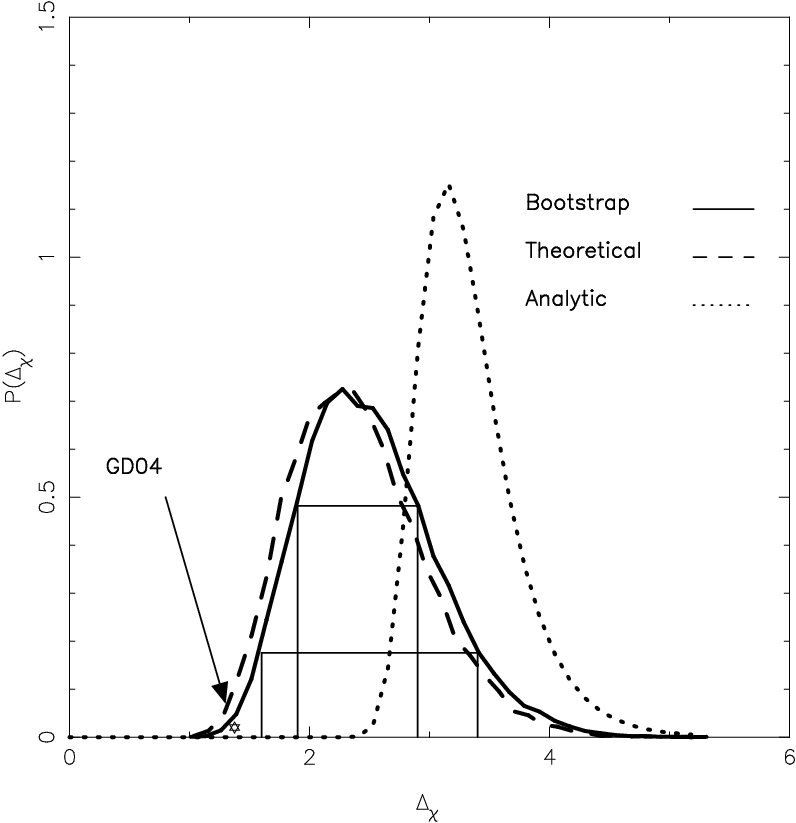}
\caption{The theoretical, analytic and the bootstrap probability distributions 
for the $\Delta_{\chi}$ statistic for GD04. The analytic
distribution uses the limiting values of the shape and position
parameters for the Gumbel distribution. Since the total number of
independent directions is most likely smaller than $N_d$, we
overestimate the position of peak. The bootstrap distribution
indicates anisotropy at the level of slightly greater than about two
sigma.  }
\label{fig:new04}
\end{figure}

{\bf GD04}: The bootstrap, theoretical  and analytic distributions  are plotted in
Fig~\ref{fig:new04}.  The analytic distribution is very different from
the other two. However, this can be explained by the fact
that our assumption of theoretical limit may not be satisfied in this
case due to a small number of SNe in the data.
 However, the bootstrap distribution agrees quite well with the
theoretical distribution thus indicating Gaussianity. Similar to the
previous statistic, we find that GD04 is slightly more than
2$\sigma$ away from the mode of the distribution. One thing to note is
that here $\Delta_{\chi}$ is smaller than the mode of the bootstrap distribution,
indicating a smaller anisotropy in the data than is expected from a
purely statistical point of view.

\begin{figure}
\centering
\includegraphics[width=0.50\textwidth]{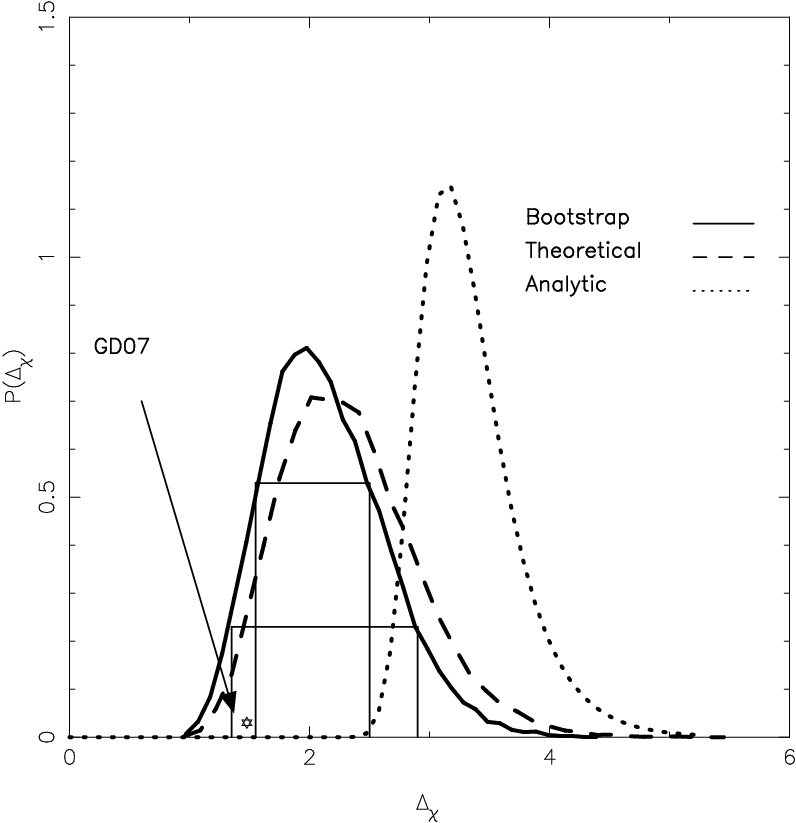}
\caption{The theoretical, analytic and the bootstrap probability distributions 
for the $\Delta_{\chi}$ statistic for GD07 as in
Figure~\ref{fig:new04}. The bootstrap distribution indicates
consistency with isotropic distribution of SNe in GD07. }
\label{fig:new07}
\end{figure}

{\bf GD07}: The results are plotted  in
Fig~\ref{fig:new07}. The main difference from GD04 is that
the anisotropy is smaller in this case. Therefore, GD07 is consistent
with Gaussianity and isotropy, in agreement with what we find with the
$\Delta_{\chi^2}$ statistic.

\section{Conclusions}

We have presented corrected results for the $\Delta$ statistic introduced in
GSL08 and showed that the results for the $\Delta_{\chi^2}$ statistic 
change substantially after correcting for 
a numerical bug in the our code. Our present method
extends the previous work by invoking an estimator that is
marginalised over the Hubble parameter. However, our results are not
affected substantially due to this change since the bootstrap
distribution and the location of $\Delta_{\chi^2}$ does not change
substantially for GD04 compared to our previous results.
However, for GD07 we have corrected another
mistake in our earlier analysis and we find that the bootstrap distribution as well
as the position of GD07 with respect to it have changed.  Our main conclusions are 1) the match between the bootstrap 
and the theoretical distribution is much better than was
presented in GSL08 2) GD04 shows some evidence for non-Gaussianity, 
however, GD07 is statistically consistent with
a Gaussian distribution of residuals 3) and although GD04 has a weak direction
dependence, GD07 is consistent with an isotropic distribution of SNe.

Our new statistic $\Delta_{\chi}$ similarly shows a weak
direction dependence in GD04 but no significant anisotropy in GD07.
One surprising feature we find is that for both GD04 and GD07,
$\Delta_{\chi^2}$ turns to be larger than the mode of the distribution, while
$\Delta_{\chi}$ is smaller than the mode of the distribution. However, since
our results are consistent with isotropy, we do not investigate the implications of 
this puzzling feature. The results of this statistic can be compared to the results
of \cite{coo09}. The directions of maximum anisotropy are different. However, since
the methods and data are different, and the results suggest only a weak direction dependence,
this mismatch is not unexpected.

\begin{table}
\begin{center}
\caption{
  The model parameters for GD04 and GD07 are tabulated  here.}  
  \label{tbl-newfit} \bigskip

\begin{tabular}{crrrrrr}
\hline
Model & Set & $\Om$ & $ H_0 $ &  $\chi^2$ \\
\hline
$\Lambda$CDM & GD04 & 0.30 & 64.5 &  177.3 \\
$\Lambda$CDM & GD07 & 0.33 & 63.0 &  158.8 \\
\hline\hline                                                                                        
\end{tabular}
\end{center}
\end{table}

\begin{table}
\begin{center}
\caption{ Direction for maximum $\Delta$ GD04 and GD07 are tabulated  here.}
  \label{tbl-newdel} \bigskip

\begin{tabular}{crrrrr}
\hline
Model & Set & $\Delta_{\chi}$ &  long  &  lat \\
\hline
$\Lambda$CDM & GD04 &-1.52 & 107 &  26 \\
$\Lambda$CDM & GD07  &-1.77  & 97 &  70 \\
\hline\hline
\end{tabular}
\end{center}
\end{table}

\appendix
\label{appendix}
\section{Marginalisation}
The normalized likelihood function for the distance modulus is given by
\begin{equation}
L(\mathbf x) = \frac{1}{(2\pi)^{N/2}\sqrt{\det {\mathbf \Sigma}}}\exp \left ( -\half \mathbf{ x}^{\T} \mathbf{\Sigma}^{-1} \mathbf{x} \right )\,.
\end{equation}
Here $x_i = \mu_i - \mu_{\rm theory} (z_i; H_0,\Om,\Ox)$, where the
subscript $i=0,\ldots,N$ and $\Sigma_{ij} = \delta_{ij}\,\sigma_i^2$ is the
covariance matrix. The superscript 'T' denotes the matrix operation of
taking the transpose. Each modulus has a linear constant term
containing a combination of parameters $H_0$ and $M$, therefore, we
wish to consider the difference of two moduli to explicitly remove
it. For concreteness we choose the $0^{\rm th}$ SN as the anchor
point, subtract its modulus from all others, and marginalise over
it. We first define these new parameters $\mathbf y$ as follows.
\begin{equation}
{\mathbf y} = \mathbf{\Lambda^{-1} x}\,,
\end{equation}
where the transformation matrix 
\[ \Lambda^{-1}_{ij} = \left\{\begin{array}{rl}
                1 &        \mbox{for $i=j$}\\
                -1&       \mbox{for $i>0, j=0$}\\
                0&        \mbox{otherwise}
                        \end{array}
        \right. \]
The inverse transformation is
\begin{equation}
\mathbf{x = \Lambda y}\,.
\label{eq:lambda2}
\end{equation}
Noting that $\det {\mathbf \Lambda} = 1$, in terms of the new
variables, the likelihood function is given by
\begin{equation}
L(\mathbf y) = \frac{1}{(2\pi)^{N/2}\sqrt{\det{{\mathbf S}}}}\exp \left ( -\half {\mathbf{ y}^{\T} \mathbf {S^{-1} y}} \right )\,\,,
\end{equation}
where ${\mathbf{S}^{-1}= \mathbf{\Lambda}^{\T}{\boldsymbol \Sigma}^{-1}\mathbf {\Lambda}}$. 

We now integrate over $y_0$ to
obtain the marginalised likelihood function,
\begin{equation}
\label{eq:marglik}
L(y_1,y_2,\ldots,y_N) = \frac{1}{(2\pi)^{N/2}\sqrt{\det{{\mathbf C}}}}\exp \left ( -\half {\mathbf{ y}^{\T} \mathbf {C}^{-1}\mathbf{ y}} \right )\,\,,
\end{equation}
where, the final covariance matrix $\mathbf C$ is given by
\begin{equation}
C^{-1}_{ij} = S^{-1}_{ij} - \frac{S^{-1}_{0i}\,S^{-1}_{0j}}{{S^{-1}_{00}}}\,\,,
\end{equation}
where the indices $i,j$ run from $1$ to $N$. To formulate our estimator for $H_0$ marginalised statistic
we require the marginalised $\chi^2_M$ defined as follows
\begin{equation}
\chi^2_M = \mathbf{ y}^{\T} \mathbf {C}^{-1}\mathbf{ y} 
\label{eq:mchisq}
\end{equation}
We note that the expectation value of $\langle \chi^2_M \rangle = N$
as shown below
\begin{equation}
\langle \chi^2_M \rangle = \langle \mathbf{ y}^{\T} \mathbf {C}^{-1}\mathbf{ y} \rangle = {\rm tr}(\mathbf {C}^{-1} \mathbf {C}) =N
\end{equation}
In this formulation the marginalisation is carried out using one of
the SN data points. However, it is easy to see that the method is more
general and we can marginalise over any Gaussian prior on the Hubble
parameter. To achieve this we note that at low redshifts the
luminosity distance takes the approximate form $ d_L \simeq
cz/H_0$. Thus, it is possible to produce a theoretical luminosity
distance at low redshifts with appropriate error bars if we are given
Gaussian priors on the Hubble parameter and the absolute magnitude for
type~Ia SNe. If we use this theoretically produced magnitude and
replace the $0^{\rm th}$ SN with it then we can marginalise over any
arbitrary priors on the Hubble parameter.


\medskip
\noindent{\it Acknowledgments:}
We thank Ryan Cooke for useful comments. 
Shashikant thanks DST (DSTO815) for providing financial assistance 
for this work.

\def\etal{{\it et~al.\ }}
\def\apj{{Astroph.\@ J.\ }}
\def\araa{{Ann. \@ Rev. \@ Astron. \@ Astroph.\ }}
\def\mn{{Mon.\@ Not.\@ Roy.\@ Ast.\@ Soc.\ }}
\def\asta{{Astron.\@ Astrophys.\ }}
\def\aj{{Astron.\@ J.\ }}
\def\prl{{Phys.\@ Rev.\@ Lett.\ }}
\def\pd{{Phys.\@ Rev.\@ D\ }}
\def\nucp{{Nucl.\@ Phys.\ }}
\def\nat{{Nature\ }}
\def\sci{{Science\ }}
\def\plb {{Phys.\@ Lett.\@ B\ }}
\def \jetpl {JETP Lett.\ }
\def \jcap {J. Cosmol. Astropart. Phys.}

\end{document}